# Student difficulties with quantum states while translating state vectors in Dirac notation to wave functions in position and momentum representations

Emily Marshman and Chandralekha Singh

*Department of Physics and Astronomy, University of Pittsburgh, Pittsburgh, PA, 15260, USA*

**Abstract.** We administered written free-response and multiple-choice questions and conducted individual interviews to investigate the difficulties that upper-level undergraduate and graduate students have with quantum states while translating state vectors in Dirac notation to wave functions in position and momentum representations. We find that students share common difficulties with translating a state vector written in Dirac notation to the wave function in position or momentum representation.

## I. INTRODUCTION

Learning quantum mechanics is challenging [1-6]. Dirac notation is commonly used in upper-level quantum mechanics. Advanced students should become facile at translating state vectors in Dirac notation to wave functions in position and momentum representations.

The generic quantum state $|\Psi\rangle$ in Dirac notation contains all information about the system. To represent the generic state $|\Psi\rangle$ as a wave function in the position representation, one must project $|\Psi\rangle$ along position eigenstates $|x\rangle$, i.e., $\langle x|\Psi\rangle$, where $x$ is a continuous index. Similarly, a generic state $|\Psi\rangle$ can be represented as a wave function in the momentum representation by projecting $|\Psi\rangle$ along momentum eigenstates $|p\rangle$, i.e., $\langle p|\Psi\rangle$, where $p$ is a continuous index. To represent position eigenstates $|x'\rangle$ with eigenvalues $x'$ and momentum eigenstates $|p'\rangle$ with eigenvalues $p'$ in the position and momentum representations, one must project them along position eigenstates $|x\rangle$ or momentum eigenstates $|p\rangle$. For example, ignoring normalization issues, a position eigenstate in the position representation is a (highly localized) delta function $\langle x|x'\rangle = \delta(x - x')$. On the other hand, a position eigenstate $|x'\rangle$ in the momentum representation is a delocalized function of momentum $\langle p|x'\rangle = e^{-ipx'/\hbar}$.

Here, we investigate student difficulties in translating a state vector in Dirac notation to a wave function in position or momentum representation. In particular, we examine the extent to which advanced students:
1) **recognize** the wave functions in the position or momentum representation written with or without Dirac notation (e.g., evaluate the correctness of the statement that a generic wave function $\Psi(x) = \langle x|\Psi\rangle$);
2) **recall** how to write wave functions in the position or momentum representation with or without Dirac notation (e.g., given $\langle x|\Psi\rangle$, recall that it can be written as $\Psi(x)$ without Dirac notation and vice versa); and
3) **generate** a wave function in position or momentum representation for (I) a generic state vector $|\Psi\rangle$, (II) position eigenstate $|x'\rangle$ with eigenvalue $x'$, or (III) momentum eigenstate $|p'\rangle$ with eigenvalue $p'$. While there are other ways to categorize these types of questions, the researchers jointly agreed that "recognize, recall, generate" is one way to code these types of questions. We discuss student difficulties with generating wave functions in position or momentum representation for different state vectors in Dirac notation.

## II. METHODOLOGY AND RESULTS

Student difficulties with these issues were investigated by administering open-ended questions and multiple-choice surveys to upper-level undergraduate and graduate students, observing common difficulties on in-class quizzes and exams, and conducting individual interviews with students enrolled in quantum mechanics courses. Open-ended questions and multiple-choice surveys were administered after traditional instruction in seven semesters of undergraduate Quantum Mechanics I courses at the University of Pittsburgh and analyzed. Multiple-choice questions were administered to upper-level students ($N = 184$) after traditional instruction at four U.S. universities.

The open-ended questions on quizzes and exams were graded using rubrics which were developed by the two investigators together. A subset of the open-ended questions was graded separately by the investigators. After comparing the grading of the open-ended questions, the investigators discussed any disagreements in grading and resolved them with a final inter-rater reliability of better than 90%.

The individual interviews employed a think-aloud protocol to better understand the rationale for student responses. During the semi-structured interviews, we asked students to "think aloud" while answering the questions. Students first read the questions on their own and answered them without interruptions except that they were prompted to think aloud if they were quiet for a long time. After students had finished answering a particular question to the best of their ability, we asked them to further clarify and elaborate issues that they had not clearly addressed earlier.

### A. Difficulties with a generic quantum state $|\Psi\rangle$

**Difficulty writing a generic state vector $|\Psi\rangle$ in position or momentum representation.** Table 1 shows that a majority of students performed well when they were asked to recognize whether a generic state vector $|\Psi\rangle$ in position representation is $\Psi(x) = \langle x|\Psi\rangle$ and in momentum representation is $\Phi(p) = \langle p|\Psi\rangle$. In particular, upper-level students ($N = 184$) were asked to evaluate the correctness of the following statement after traditional instruction given a generic state vector $|\Psi\rangle$: *"The wave function in position representation is $\Psi(x) = \langle x|\Psi\rangle$ where $x$ is a continuous index."* Eighty-nine percent of the students agreed with this statement, indicating that they recognize that the wave function in position representation is $\Psi(x) = \langle x|\Psi\rangle$. The same students also evaluated the correctness of the following statement: *"The wave function in momentum representation is $\Phi(p) = \langle p|\Psi\rangle$ where $p$ is a continuous index."* Table I shows that 77% of the students agreed with this statement, which indicates that they correctly recognize that the wave function in momentum representation is $\Phi(p) = \langle p|\Psi\rangle$ (although this percentage is smaller than the percentage of students who correctly recognized $\Psi(x) = \langle x|\Psi\rangle$).

Table I also shows that when upper-level students ($N = 127$) were asked to describe the physical significance of $\langle x|\Psi\rangle$ on a midterm exam after traditional instruction, 86% of them correctly recalled that $\langle x|\Psi\rangle = \Psi(x)$ and that $\Psi(x)$ is also known as the wave function in position representation (some even related it to the probability density for measuring position). Similarly, when these students were asked to describe the physical significance of $\langle p|\Psi\rangle$ on the same midterm exam, 85% of them correctly recalled that $\langle p|\Psi\rangle = \Phi(p)$ and $\Phi(p)$ is known as the wave function in momentum representation (see Table I).

In contrast, Table I shows that students had difficulty generating on their own how to write a generic state vector $|\Psi\rangle$ in the position representation. For example, 46 upper-level students were asked the following question after traditional instruction: *You are given a generic state vector $|\Psi\rangle$. How would you obtain the wave function in position representation from $|\Psi\rangle$?* Answers were considered correct if students wrote $\langle x|\Psi\rangle$, $\Psi(x)$, or stated that one needs to project the generic state $|\Psi\rangle$ onto the position basis, i.e., $\langle x|\Psi\rangle = \Psi(x)$. Only 52% provided the correct response.

These types of responses indicate that students may be adept at recognizing and recalling answers to questions about translating a generic state vector between Dirac notation and position and momentum representations. However, many struggle to generate the wave function in the position and momentum representations given state vector $|\Psi\rangle$. In other words, depending on the cues or scaffolding provided in the problem statement (e.g., whether the question asked is in the recognize, recall, or generate category), students may have different levels of difficulty in translating a state vector from Dirac notation to position and momentum representations. The difference in the difficulty level in recognizing, recalling and generating indicates that students are still developing expertise and their knowledge structure is not robust [7].

**Confusing a state with an operator in the context of a generic state vector $|\Psi\rangle$.** As noted, many students had difficulty generating the wave function in position representation given the generic state vector $|\Psi\rangle$. One of the most common difficulties was generating a response which included the position operator. Table II shows that of the 46 upper-level students, 28% provided responses which involved the position operator. Common incorrect responses of this type included, e.g., $\hat{x}|\Psi\rangle = x|\Psi\rangle$, $\hat{x}|\Psi\rangle = \langle x|\Psi\rangle$, $\langle x|\Psi\rangle = \int \hat{x}^*\Psi dx = \int x\,\Psi dx$, and $\hat{x}\,\Psi(x) = \langle x|\hat{x}|\Psi\rangle$. Students displayed similar difficulties with a generic state $|\Psi\rangle$ in momentum representation, with common incorrect responses of the form $\langle p|\Psi\rangle = \int \hat{p}^*\Psi dx = \int i\hbar\,\partial/\partial x\,\Psi dx$.

**Table I**. Percentages of undergraduate students who correctly answered questions related to writing the quantum state $|\Psi\rangle$ in position representation (N = number of students).

| | |
|---|---|
| **Recognize:** Evaluate the correctness of the statement: "The wave function in position representation is $\Psi(x) = \langle x|\Psi\rangle$ where $x$ is a continuous index" ($N = 184$) | 89 |
| **Recognize:** Evaluate the correctness of the statement: "The wave function in momentum representation is $\Phi(p) = \langle p|\Psi\rangle$ where $p$ is a continuous index." ($N = 184$) | 77 |
| **Recall:** What is the physical significance of $\langle x|\Psi\rangle$? ($N = 127$) | 86 |
| **Recall:** What is the physical significance of $\langle p|\Psi\rangle$? ($N = 127$) | 85 |
| **Generate:** How would you obtain the wave function in position representation from $|\Psi\rangle$? ($N = 46$) | 52 |

Furthermore, although many upper-level students ($N = 127$) correctly recalled that $\langle x|\Psi\rangle = \Psi(x)$ and $\Psi(x)$ is also known as the wave function in position representation when asked to describe the physical significance of $\langle x|\Psi\rangle$, they often wrote additional incorrect statements in their responses claiming that the position (or momentum) operator is involved in determining the wave function in the position or momentum representation. Table II shows that 25% of the students claimed that the position (or momentum) operator is involved in determining the wave function in the position (or momentum) representation. For example, one student stated that "$\langle x|\Psi\rangle$ is just $\int x^*\Psi dx = \Psi$ in position basis." Another student incorrectly claimed that "$\langle x|\Psi\rangle$ is the measurement of $|\Psi\rangle$ in position, it yields a position eigenstate of the system at the time of measurement." Similarly, in response to the question about the physical significance of $\langle p|\Psi\rangle$, another student stated "$\langle p|\Psi\rangle = \int \Psi\left(-\frac{\hbar}{i}\frac{\partial}{\partial x}\right)\Psi dx = \frac{\hbar}{i}\int \Psi\left(\frac{\partial}{\partial x}\right)\Psi$." These types of responses indicate that students have difficulty distinguishing between the projection of a state vector $|\Psi\rangle$ along an eigenstate of $x$ or $p$ vs. the position or momentum operator acting on a generic state vector $|\Psi\rangle$. These responses also suggest that students have difficulty with the physical significance of $\langle x|\Psi\rangle$ or $\langle p|\Psi\rangle$, which are the probability density amplitudes for measuring $x$ or $p$.

When students were asked to evaluate the correctness of a statement in which this type of difficulty is explicitly mentioned (e.g., confusion between representing a generic state in position or momentum representation by projecting a state along an eigenstate of $x$ or $p$ vs. operating on a state with the position or momentum operator), a larger percentage of the students display this type of a difficulty.

Table II shows that even when students ($N = 184$) were asked to evaluate the correctness of the statement connecting the wave function in position and momentum representation: *"The wavefunction in momentum representation is $\Phi(p) = \int dx(-i\hbar \frac{\partial}{\partial x}\Psi(x))$,"* 61% incorrectly agreed with this statement, indicating that they thought the momentum operator written in the position representation (i.e., $-i\hbar\frac{\partial}{\partial x}$) connects the wave function in momentum and position representations. This response is incorrect because one must use a Fourier transform to obtain $\Phi(p)$ from $\Psi(x)$, i.e., $\Phi(p) = \langle p|\Psi\rangle = \int dx \langle p|x\rangle\langle x|\Psi\rangle = \int dx e^{-ipx'/\hbar}\Psi(x)$.

### B. Difficulties in representing $|x'\rangle$ and $|p'\rangle$ in the position or momentum representation

In addition to exhibiting difficulties with writing a generic state $|\Psi\rangle$ in position and momentum representations, students also struggled to translate position and momentum eigenstates from Dirac notation to the position and momentum representations. Many undergraduate students struggled to recall how to write $\langle x|x'\rangle$, $\langle p|x'\rangle$, $\langle p|p'\rangle$, and $\langle x|p'\rangle$ without using Dirac notation as a function of position or momentum.

**Confusing a state with an operator in the context of position or momentum eigenstates:** Similar to the difficulty involving confusion between a bra state and an operator in the context of a generic state $|\Psi\rangle$, students confuse a state with an operator in the context of position or momentum eigenstates. For example, in determining $\langle x|x'\rangle$ in position representation without using Dirac notation, students often treated the bra state $\langle x|$ as $\hat{x}$ and incorrectly acted with it on the eigenstate $|x'\rangle$. Table II shows that when upper-level students ($N = 46$) were asked to write $\langle x|x'\rangle$ without Dirac notation as a function of $x$ after traditional instruction, one difficulty was writing $\langle x|x'\rangle = x'$, which was displayed by 13% of the students. Interviews suggest that this type of difficulty sometimes stemmed from the fact that students treated the bra state $\langle x|$ as the position operator $\hat{x}$ and acted with it on $|x'\rangle$ and then incorrectly removed the state $|x'\rangle$ after the operation, e.g., $\langle x|x'\rangle = \hat{x}|x'\rangle = x'$.

Similar difficulties are displayed when 46 upper-level students were asked to write $\langle p|p'\rangle$ without Dirac notation after instruction in relevant concepts. Table II shows that one difficulty was writing $\langle p|p'\rangle = p'$, which was displayed by 9% of the students. Interviews suggest that this type of difficulty sometimes stemmed from the fact that students treated the bra state $\langle p|$ as the momentum operator $\hat{p}$ and acted with it on $|p'\rangle$ and then incorrectly removed the state $|p'\rangle$ after the operation, e.g., $\langle p|p'\rangle = \hat{p}|p'\rangle = p'$

**Assuming $\langle x|x'\rangle = 1$ or $0$ (or $\langle p|p'\rangle = 1$ or $0$):** Upper-level students ($N = 46$) were asked to write $\langle x|x'\rangle$ and $\langle p|p'\rangle$ without using Dirac notation after traditional instruction. Responses were considered correct if the students wrote $\langle x|x'\rangle = \delta(x - x')$ (or $\langle p|p'\rangle = \delta(p - p')$). Table III shows that only 35% answered correctly. Some students incorrectly invoked a "normalization condition" when determining $\langle x|x'\rangle$ or $\langle p|p'\rangle$. For example, Table II shows that 6% wrote that $\langle x|x'\rangle = 1$ and 7% wrote that $\langle p|p'\rangle = 1$. In interviews, students often incorrectly claimed that $\langle x|x'\rangle = 1$ if $x = x'$ or $\langle x|x'\rangle = 0$ if $x \neq x'$. Further discussion with students suggests that this type of difficulty was often the result of confusing the Kronecker delta and the Dirac delta function. The Kronecker delta is appropriate to use, e.g., for orthogonality of eigenstates with discrete eigenvalues (i.e., $\delta_{nm} = 1$ if $n = m$ and $\delta_{nm} = 0$ if $n \neq m$). The Dirac delta function, e.g., $\delta(x - x')$, is appropriate for eigenstates with continuous eigenvalues. When $x = x'$, $\delta(x - x')$ is infinite.

**Table II.** Percentages of students displaying difficulties with quantum states in position and momentum representations.

| Difficulty | Question statement | % |
|---|---|---|
| Confusing a state with an operator in the context of a generic state vector $|\Psi\rangle$ | You are given a generic state vector $|\Psi\rangle$. How would you obtain the wave function in position representation from $|\Psi\rangle$? ($N = 46$) | 28 |
| | What is the physical significance of $\langle x|\Psi\rangle$? ($N = 127$) | 25 |
| | Evaluate the correctness of the statement: "The wave function in momentum representation is $\Psi(p) = \int dx(-i\hbar\frac{\partial}{\partial x}\Psi(x))$." ($N = 184$) | 61 |
| Confusing a state with an operator in the context of position or momentum eigenstates $|x'\rangle$ or $|p'\rangle$ | $\langle x|x'\rangle = ?$  ($N = 46$) | 13 |
| | $\langle p|p'\rangle = ?$  ($N = 46$) | 9 |
| Assuming $\langle x|x'\rangle = 1$ or $0$ (or $\langle p|p'\rangle = 1$ or $0$) | $\langle x|x'\rangle = ?$  ($N = 46$) | 6 |
| | $\langle p|p'\rangle = ?$  ($N = 46$) | 7 |
| Assuming $\langle x|p'\rangle = 0$ or $\langle p|x'\rangle = 0$ | $\langle x|p'\rangle = ?$  ($N = 46$) | 9 |
| | $\langle p|x'\rangle = ?$  ($N = 46$) | 9 |

**Assuming $\langle x|p'\rangle = 0$ or $\langle p|x'\rangle = 0$:** Forty-six upper-level students were also asked to write $\langle p|x'\rangle$ and $\langle x|p'\rangle$ without Dirac notation as a function of position or momentum after instruction in relevant concepts. Table III shows that 20% correctly recalled that $\langle p|x'\rangle = e^{-ipx'/\hbar}$ and $\langle x|p'\rangle = e^{ip'x/\hbar}$. Students were not penalized if they did not write down a constant pre-factor often used as "normalization" or if they did not have the correct sign in the exponent. A common difficulty involved invoking an orthogonality condition. For example, Table II shows that 9% of the students wrote $\langle p|x'\rangle = 0$ or $\langle x|p'\rangle = 0$. In interviews, some students who had traditional instruction in these issues initially stated that eigenstates of $x$ and $p$ are orthogonal. Others stated that since $x$ and $p$ were incompatible, the inner products $\langle p|x'\rangle$ or $\langle x|p'\rangle$ did not make sense. Students stated that if it was appropriate to have such inner products, they must be zero because $x$ and $p$ have "nothing in common." Prior research shows that even in the context of a two-dimensional vector space for a spin-1/2 system, students often make similar claim, e.g., that eigenstates of $\hat{S}_x$ are orthogonal to eigenstates of $\hat{S}_y$ [6].

### C. Performance of graduate students

**Graduate students perform significantly better on questions involving recall:** Graduate students enrolled in a first year core graduate quantum mechanics course were more proficient than undergraduates at translating between Dirac notation and position and momentum representations. For example, Table III shows that 45 graduate students, who were asked to write $\langle x|x'\rangle, \langle x|p'\rangle, \langle p|x'\rangle$ and $\langle p|p'\rangle$ in position and momentum representation without using Dirac notation, performed significantly better on average than the undergraduate students.

**Graduate students have difficulty in generating answers to questions on these topics although they are good at recall:** Table III shows that when 45 graduate students were asked to write, e.g., $\langle p|p'\rangle$ or $\langle x|p'\rangle$ without Dirac notation, 91% correctly wrote, e.g., that $\langle p|p'\rangle = \delta(p - p')$ which is significantly higher than 35%, the corresponding average undergraduate percentage. However, only 49% of the graduate students correctly answered the question "write a momentum eigenstate with eigenvalue $p'$ in momentum representation." Responses were considered correct if the student wrote $\langle p|p'\rangle$ or $\delta(p - p')$. What is noteworthy is that on the same survey, 42% of the graduate students correctly recalled but could not generate a related answer, e.g., they correctly wrote $\langle p|p'\rangle = \delta(p - p')$ but answered incorrectly when asked to generate a momentum eigenstate with eigenvalue $p'$ in the momentum representation. Similarly, 29% of the graduate students correctly answered questions asking them to recall a momentum eigenstate with eigenvalue $p'$ in position representation but could not generate it, i.e., they correctly recalled $\langle x|p'\rangle = e^{ip'x/\hbar}$ but answered incorrectly when asked to generate a momentum eigenstate with eigenvalue $p'$ in position representation.

This type of dichotomy between recall vs. generate questions shown in Table III suggests that while most graduate students are proficient at recalling how to convert expressions written in Dirac notation to a form without Dirac notation, almost half of them do not understand the physical meaning of those expressions. The task of generating a momentum eigenstate with eigenvalue $p'$ in momentum or position representation requires understanding of the symbols in Dirac notation and position or momentum representation. If graduate instruction only focuses on problem solving requiring recall of these types of expressions from what was discussed in a particular context and reproducing them on the exams, students are unlikely to develop a functional understanding of these expressions.

**Table III.** Percentages of undergraduates (UG) ($N = 46$) and graduate students (G) ($N = 45$) who correctly answered questions related to position and momentum representations.

| Question | UG | G |
|---|---|---|
| **Recall:** $\langle x|x'\rangle = ?$ | 35 | 91 |
| **Recall:** $\langle p|x'\rangle = ?$ | 20 | 82 |
| **Recall:** $\langle x|p'\rangle = ?$ | 20 | 87 |
| **Recall:** $\langle p|p'\rangle = ?$ | 35 | 91 |
| **Generate:** "write a momentum eigenstate with eigenvalue $p'$ in position representation." | 13 | 58 |
| **Generate:** "write a momentum eigenstate with eigenvalue $p'$ in momentum representation." | 15 | 49 |

### III. SUMMARY

Translating between representations is a hallmark of expertise and is important for developing expertise in quantum mechanics. After traditional instruction in relevant concepts, undergraduates and even graduate students, who are proficient at recalling how to write an expression given in Dirac notation without the use of the Dirac notation (or vice versa), have difficulties in generating their own solutions, e.g., when asked to write the position or momentum eigenstates in position and momentum representations. Students must be given multiple opportunities to not only recognize and recall but also generate answers to these types of questions related to translation between the representations discussed here using research-based learning tools to develop a functional understanding of the underlying concepts.

### ACKNOWLEDGEMENTS

We thank the National Science Foundation for award PHY-1202909.